\documentclass{article}
\usepackage{arxiv}
\usepackage[utf8]{inputenc}
\usepackage[T1]{fontenc}
\usepackage{amsmath, amsfonts, amssymb, amsthm}
\usepackage{graphicx}
\usepackage{booktabs}
\usepackage{hyperref}
\usepackage{algorithm}
\usepackage{algpseudocode}
\usepackage{xcolor}
\usepackage{listings}
\usepackage{pdflscape} 

\definecolor{commentcolor}{rgb}{0.5,0.5,0.5}
\definecolor{keywordcolor}{rgb}{0.1,0.1,0.7}
\definecolor{stringcolor}{rgb}{0.1,0.6,0.1}

\usepackage{bm}
\usepackage{multirow}

\newtheorem{remark}{Remark}

\title{Parametric modal regression for right-censored positive responses}

\author{
  Christian E.\ Galarza \\
  Escuela Superior Polit\'ecnica del Litoral\\
  Guayaquil, Ecuador \\
  \texttt{chedgala@espol.edu.ec}
  \And
  V\'ictor H.\ Lachos \\
  University of Connecticut \\
  Storrs, CT, USA \\
  \texttt{hlachos@uconn.edu}
}

\begin{document}
\maketitle

\begin{abstract}
We present a unified parametric framework for modal regression applicable to continuous positive distributions, with explicit support for right-censored observations. The key contribution is a systematic analytical reparameterization of density parameters as direct functions of the conditional mode. This closed-form mapping is derived for the Gamma, Beta, Weibull, Lognormal, and Inverse Gaussian distributions, directly linking the mode to a linear predictor. Maximum likelihood estimation is performed using the censored log-likelihood, with asymptotic inference based on the observed Fisher information matrix. A Monte Carlo simulation study across multiple distributions, sample sizes, and censoring levels confirms consistent parameter recovery. Empirical bias and RMSE decrease as expected, and Wald confidence intervals achieve nominal coverage. Finally, the proposed methodology is illustrated through an application to real-world reliability data. All methodology is implemented in the open-source R package \texttt{ModalCens}.
\end{abstract}

\textbf{Keywords:} modal regression; exponential family; right censoring;
maximum likelihood; randomized quantile residuals; R package.

\section{Introduction}
\label{sec:intro}

Classical Generalized Linear Models (GLMs; \cite{mccullagh1989}) target
the conditional mean $\mathbb{E}[Y_i \mid \bm{x}_i]$, which provides an
optimal summary of location when the response distribution is symmetric or
moderately skewed. For distributions with positive support and pronounced
right skewness---such as those arising in survival analysis, reliability
engineering, and environmental monitoring---the conditional mean is
sensitive to extreme observations in the upper tail and may not correspond
to any high-density region of the response. In such contexts, the
conditional \emph{mode} $M_i = \arg\max_{y} f(y \mid \bm{x}_i)$ is a more
interpretable and robust characterization of the ``most likely'' outcome.

Modal regression---modeling $M_i$ as a function of covariates---has
attracted increasing attention in the nonparametric literature
\cite{lee1989,yao2014,chen2016,xiang2026}. The semiparametric estimator of
\cite{yao2014} exploits kernel density estimation of the residual
distribution to define a modal regression objective without specifying a
parametric family. While flexible, this approach does not naturally extend
to censored settings, and the absence of a parametric likelihood precludes
standard information-based inference and model comparison criteria such as
AIC and BIC.

The present work adopts a \emph{strictly parametric} approach. We
reparameterize common continuous positive distributions so that the
conditional mode $M_i$ appears as the structural parameter, replacing the
canonical parameterization. For distributions belonging to the exponential
family, the reparameterization follows from the canonical form; for other
families, the same first-order optimality condition of the log-density is
applied directly. In both cases the mapping is exact and yields
a fully standard maximum likelihood problem in which inference, model
selection, and residual diagnostics follow from classical theory. Right
censoring is accommodated by the standard likelihood factorization that
replaces the density contribution of a censored observation with its
survival function.

The resulting methodology is implemented in the R package \texttt{ModalCens},
which supports five distribution families (Gamma, Beta, Weibull, Lognormal, and Inverse
Gaussian), provides a formula-based interface consistent with standard R
modeling conventions, and exports S3 methods for \texttt{summary},
\texttt{plot}, \texttt{AIC}, \texttt{BIC}, \texttt{coef}, \texttt{vcov},
\texttt{residuals}, and \texttt{logLik}.

The remainder of the paper is organized as follows. Section~\ref{sec:framework}
establishes the general modal parameterization framework. Section~\ref{sec:cases}
derives the mode reparameterizations for each supported family.
Section~\ref{sec:censlik} formulates the censored log-likelihood and
the estimation algorithm. Section~\ref{sec:inference} discusses asymptotic
inference and diagnostic residuals. Section~\ref{sec:package} documents
the \texttt{ModalCens} package. Section~\ref{sec:simulation} reports a
comprehensive Monte Carlo simulation study evaluating parameter recovery,
RMSE, and coverage probability. Section~\ref{sec:application} illustrates
the methodology on a real dataset. Section~\ref{sec:conclusion} concludes
and outlines future research directions.

\section{General Framework: Modal Parameterization}
\label{sec:framework}

\subsection{Motivating Case: The Exponential Family}

Let $Y_i$ be a continuous positive (or positively bounded) random variable
whose density belongs to the exponential family:
\begin{equation}
  f(y_i \mid \theta_i, \phi)
  = \exp\!\left\{
      \frac{y_i \theta_i - b(\theta_i)}{a(\phi)} + c(y_i, \phi)
    \right\},
  \quad y_i \in \mathcal{Y} \subseteq (0, \infty),
  \label{eq:expfam}
\end{equation}
where $\theta_i$ is the canonical parameter, $\phi > 0$ is the dispersion
parameter, and $b(\cdot)$, $a(\cdot)$, $c(\cdot, \cdot)$ are
family-specific functions satisfying the standard regularity conditions
that ensure $\mathbb{E}[Y_i] = b'(\theta_i)$ and
$\mathrm{Var}(Y_i) = a(\phi)\, b''(\theta_i)$.

For a strictly log-concave density whose mode lies in the interior of
$\mathcal{Y}$, the conditional mode $M_i$ satisfies the first-order
stationarity condition
\begin{equation}
  \left.\frac{\partial \log f(y_i \mid \theta_i, \phi)}{\partial y_i}
  \right|_{y_i = M_i} = 0,
  \label{eq:foc}
\end{equation}
which, from \eqref{eq:expfam}, expands to
\begin{equation}
  \frac{\theta_i}{a(\phi)} + c'(M_i, \phi) = 0,
  \label{eq:foc2}
\end{equation}
where $c'(y_i, \phi) = \partial c(y_i, \phi) / \partial y_i$. Solving
\eqref{eq:foc2} for $\theta_i$ yields the \emph{analytical mode mapping}:
\begin{equation}
  \theta_i(M_i, \phi) = -a(\phi)\, c'(M_i, \phi).
  \label{eq:mapping}
\end{equation}
This mapping is exact: for any interior mode $M_i$, equation
\eqref{eq:mapping} identifies the canonical parameter that places the
mode of the density precisely at $M_i$. Substituting \eqref{eq:mapping}
back into \eqref{eq:expfam} yields a density whose natural parameter is
$M_i$, enabling direct modal regression. Among the five families
considered in this paper, the Gamma, Beta, and Inverse Gaussian
distributions belong to the exponential family; their mode
reparameterizations follow directly from \eqref{eq:mapping} and are
derived in Sections~\ref{sec:gamma}, \ref{sec:beta},
and~\ref{sec:ig}.

\subsection{Extension to Non-Exponential Family Distributions}

The key insight underlying \eqref{eq:mapping}---that the mode can be
expressed as a function of the density parameters by solving the first-order
condition $\partial \log f / \partial y = 0$---does not require the
exponential family structure. For any continuous positive distribution with
a differentiable log-density and a unique interior mode, the stationarity
condition \eqref{eq:foc} provides an implicit equation relating the mode
$M_i$ to the distributional parameters. Inverting this equation yields an
analogous mode reparameterization.

In particular, the Weibull and Lognormal distributions---which do not
belong to the exponential family when their shape or scale parameters are
unknown---admit closed-form mode mappings derived from the same first-order
principle. Their reparameterizations are presented in
Sections~\ref{sec:weibull} and~\ref{sec:lognormal}.

\begin{remark}
The mode reparameterization---whether obtained via \eqref{eq:mapping} for
exponential family members or via the general stationarity condition
\eqref{eq:foc} for other families---requires that the log-density is
strictly concave in $y_i$ so that the stationarity condition has a unique
interior solution. This holds for all five families considered below.
Distributions whose density is strictly decreasing on
$(0,\infty)$---e.g., the Exponential---do not possess an interior mode
and therefore cannot be modal-regressed via this framework.
\end{remark}

\subsection{Modal Regression Model}

Let $\bm{x}_i \in \mathbb{R}^p$ denote the covariate vector for unit $i$.
The modal regression model specifies
\begin{equation}
  g(M_i) = \bm{x}_i^\top \bm{\gamma},
  \label{eq:link}
\end{equation}
where $g(\cdot): \mathcal{Y} \to \mathbb{R}$ is a monotone, differentiable
link function and $\bm{\gamma} \in \mathbb{R}^p$ is the vector of regression
coefficients. The full parameter vector is $\bm{\Theta} = (\bm{\gamma}^\top, \phi)^\top$.
Table~\ref{tab:links} summarizes the link functions used for each supported
family.

\begin{table}[h]
  \caption{Mode domain and link functions for the supported distributions.
           $^\dagger$The Inverse Gaussian reparameterization requires
           $\lambda > 3M_i$ at every observation; see Section~\ref{sec:cases}.}
  \label{tab:links}
  \centering
  \begin{tabular}{lccc}
    \toprule
    Distribution & Mode Domain & Link function & Mode Model \\
    \midrule
    Gamma         & $(0,\infty)$ & $\log$  & $\exp(\bm{x}_i^\top\bm{\gamma})$ \\
    Beta          & $(0,1)$      & $\mathrm{logit}$ & $\exp(\bm{x}_i^\top\bm{\gamma}) [1+\exp(\bm{x}_i^\top\bm{\gamma})]^{-1}$ \\
    Weibull       & $(0,\infty)$ & $\log$  & $\exp(\bm{x}_i^\top\bm{\gamma})$ \\
    Lognormal     & $(0,\infty)$ & $\log$  & $\exp(\bm{x}_i^\top\bm{\gamma})$ \\
    Inverse Gaussian$^\dagger$ & $(0,\infty)$ & $\log$ & $\exp(\bm{x}_i^\top\bm{\gamma})$ \\
    \bottomrule
\end{tabular}
\end{table}

\section{Mode reparameterizations for specific families}
\label{sec:cases}

\subsection{Gamma distribution}
\label{sec:gamma}

For $Y_i \sim \mathrm{Gamma}(\alpha, \beta_i)$ with shape $\alpha > 1$ and
scale $\beta_i$, the density is
$f(y_i) = y_i^{\alpha-1} e^{-y_i/\beta_i} / [\beta_i^\alpha \Gamma(\alpha)]$.
The canonical parameter is $\theta_i = -1/\beta_i$ and $a(\phi) = 1$,
so $c'(y_i, \alpha) = (\alpha-1)/y_i$. Applying \eqref{eq:mapping}:
\begin{equation}
  -\frac{1}{\beta_i} = -\frac{\alpha-1}{M_i}
  \quad\Longrightarrow\quad
  \beta_i = \frac{M_i}{\alpha - 1}.
  \label{eq:gamma_map}
\end{equation}
An alternative convention, equivalent to setting $\phi = \alpha^{-1}$
(the index of dispersion), gives $\alpha = \phi^{-1} + 1$ and
$\beta_i = M_i \phi$. The log-likelihood is
\begin{equation}
  \ell(\bm{\gamma}, \alpha) = \sum_{i=1}^n
  \left[
    (\alpha-1)\log y_i
    - \frac{y_i(\alpha-1)}{\exp(\bm{x}_i^\top\bm{\gamma})}
    - \alpha \log\!\left(\frac{\exp(\bm{x}_i^\top\bm{\gamma})}{\alpha-1}\right)
    - \log\Gamma(\alpha)
  \right].
  \label{eq:loglik_gamma}
\end{equation}

\subsection{Beta distribution}
\label{sec:beta}

For $Y_i \sim \mathrm{Beta}(\alpha, \beta_{i})$ with $\alpha, \beta_{i} > 1$
(ensuring an interior mode), the first-order condition \eqref{eq:foc} gives
\begin{equation}
  \frac{\alpha - 1}{M_i} - \frac{\beta_{i} - 1}{1 - M_i} = 0
  \quad\Longrightarrow\quad
  M_i = \frac{\alpha - 1}{\alpha + \beta_{i} - 2}.
  \label{eq:beta_mode}
\end{equation}
Fixing $\alpha$ as a global shape parameter and isolating $\beta_{i}$
as a function of the conditional mode:
\begin{equation}
  \beta_{i} = \frac{\alpha - 1}{M_i} - \alpha + 2.
  \label{eq:beta_map}
\end{equation}
With the logit link, $M_i = \exp(\eta_i)/(1+\exp(\eta_i))$ where
$\eta_i = \bm{x}_i^\top\bm{\gamma}$, and $\beta_{i}$ inherits
individual variation through $M_i$.

\subsection{Weibull Distribution}
\label{sec:weibull}

The Weibull distribution does not belong to the exponential family when the
shape parameter $k$ is unknown. Nevertheless, the mode reparameterization
follows from the same first-order condition \eqref{eq:foc} applied directly
to the log-density.

For $Y_i \sim \mathrm{Weibull}(k, \lambda_i)$ with shape $k > 1$ and
scale $\lambda_i$, the mode is $M_i = \lambda_i[(k-1)/k]^{1/k}$.
Inverting gives the scale reparameterization:
\begin{equation}
  \lambda_i = M_i \left(\frac{k}{k-1}\right)^{1/k}.
  \label{eq:weibull_map}
\end{equation}
With $k = \phi + 1.01$ (a fixed offset that enforces $k > 1$ for all
$\phi \geq 0$, ensuring the existence of an interior mode; the value $1.01$
rather than $1$ provides a small buffer that prevents numerical
instabilities near the boundary), the log-likelihood simplifies to
\begin{equation}
  \ell(\bm{\gamma}, k) = \sum_{i=1}^n
  \left[
    \log k + (k-1)\log y_i - k(\bm{x}_i^\top\bm{\gamma})
    - \log\!\left(\frac{k}{k-1}\right)
    - \left(\frac{k-1}{k}\right)
      \left(\frac{y_i}{\exp(\bm{x}_i^\top\bm{\gamma})}\right)^k
  \right].
  \label{eq:loglik_weibull}
\end{equation}

\subsection{Lognormal Distribution}
\label{sec:lognormal}

As with the Weibull, the Lognormal distribution does not belong to the
exponential family when $\sigma^2$ is unknown; the mode reparameterization
is obtained by solving the first-order condition \eqref{eq:foc} directly.

For $Y_i \sim \mathrm{Lognormal}(\mu_i, \sigma^2)$ with location parameter $\mu_i$ and scale $\sigma > 0$, the mode is given by $M_i = \exp(\mu_i - \sigma^2)$.
Inverting this relationship isolates the location parameter $\mu_i$ as a function of the conditional mode and the dispersion parameter:
\begin{equation}
  \mu_i = \log(M_i) + \sigma^2.
  \label{eq:lognormal_map}
\end{equation}
Using the logarithmic link $\log(M_i) = \bm{x}_i^\top\bm{\gamma}$ and defining the variance as a global dispersion parameter $\sigma^2 = \phi$, the log-likelihood function simplifies to
\begin{equation}
  \ell(\bm{\gamma}, \phi) = \sum_{i=1}^n
  \left[
    -\frac{1}{2}\log(2\pi\phi) - \log y_i
    - \frac{\left(\log y_i - \bm{x}_i^\top\bm{\gamma} - \phi\right)^2}{2\phi}
  \right].
  \label{eq:loglik_lognormal}
\end{equation}

\subsection{Inverse Gaussian distribution}
\label{sec:ig}

For $Y_i \sim \mathrm{IG}(\mu_i, \lambda)$ with mean $\mu_i > 0$ and shape
$\lambda > 0$, the density is
$f(y_i) = \sqrt{\lambda/(2\pi y_i^3)}\exp[-\lambda(y_i-\mu_i)^2/(2\mu_i^2 y_i)]$.
The canonical parameter is $\theta_i = -\lambda/(2\mu_i^2)$, and
applying \eqref{eq:mapping} yields
\begin{equation}
  \frac{1}{\mu_i^2} = \frac{1}{M_i^2} - \frac{3}{\lambda M_i}
  \quad\Longrightarrow\quad
  \mu_i = \left(\frac{1}{M_i^2} - \frac{3}{\lambda M_i}\right)^{-1/2},
  \label{eq:ig_map}
\end{equation}
subject to the parametric restriction $\lambda > 3M_i$, which ensures that
$\mu_i$ is real and positive. This restriction places an implicit lower
bound on $\lambda$ that depends on the fitted mode and must be monitored
during optimization.

\section{Censored Likelihood and Estimation}
\label{sec:censlik}

\subsection{Right-Censored Log-Likelihood}

Let $(y_i, c_i)$, $i = 1, \ldots, n$, be the observed data, where
$y_i$ is the observed time (exact if $c_i = 0$, right-censored if
$c_i = 1$). Under non-informative censoring, the log-likelihood is
\begin{equation}
  \ell(\bm{\gamma}, \phi)
  = \sum_{i=1}^n
  \left[
    (1-c_i)\log f(y_i \mid M_i, \phi)
    + c_i \log S(y_i \mid M_i, \phi)
  \right],
  \label{eq:censloglik}
\end{equation}
where $S(\cdot) = 1 - F(\cdot)$ is the survival function and $M_i = g^{-1}(\bm{x}_i^\top\bm{\gamma})$
is the family-specific inverse link. For each family, both $f(\cdot)$ and
$S(\cdot)$ are evaluated at the reparameterized parameters derived in
Section~\ref{sec:cases}. Note that $\ell$ reduces to the complete-data
log-likelihood when $c_i = 0$ for all $i$.

\subsection{Maximum Likelihood Estimation via BFGS}

Because the score equations $\nabla_{\bm{\Theta}} \ell = \bm{0}$ do not
admit closed-form solutions due to the transcendental nature of the
mappings in Section~\ref{sec:cases}, estimation is performed numerically.
The dispersion parameter $\phi$ is transformed to the unconstrained scale
via $\phi = \exp(\psi)$, so that the optimization variable is
$\bm{\Theta}^\star = (\bm{\gamma}^\top, \psi)^\top \in \mathbb{R}^{p+1}$.
Algorithm~\ref{alg:mle} summarizes the procedure.

\begin{algorithm}[h]
\caption{Maximum Likelihood Estimation for Modal Regression}
\label{alg:mle}
\begin{algorithmic}[1]
\Require{$\bm{y}$, $\bm{c}$, $\mathbf{X}$, family, tolerance $\varepsilon$}
\State \textbf{Initialize:} $\hat{\bm{\gamma}}^{(0)} \leftarrow$ OLS estimates of $g(y_i)$ on $\mathbf{X}$; $\hat{\psi}^{(0)} \leftarrow 0$
\State Set $\bm{\Theta}^{\star(0)} = \bigl(\hat{\bm{\gamma}}^{(0)\top}, \hat{\psi}^{(0)}\bigr)^\top$
\Repeat
  \State Compute $M_i^{(t)} = g^{-1}\!\left(\bm{x}_i^\top \hat{\bm{\gamma}}^{(t)}\right)$ and $\phi^{(t)} = \exp(\hat{\psi}^{(t)})$
  \State Evaluate $\ell\!\left(\bm{\Theta}^{\star(t)}\right)$ using \eqref{eq:censloglik} with family-specific $f$ and $S$
  \State Compute (or approximate) Score $\mathbf{U}^{(t)} = \nabla_{\bm{\Theta}^\star}\ell$ and Hessian $\tilde{\mathbf{H}}^{(t)}$
  \State Update: $\bm{\Theta}^{\star(t+1)} \leftarrow \bm{\Theta}^{\star(t)} - \left[\tilde{\mathbf{H}}^{(t)}\right]^{-1} \mathbf{U}^{(t)}$ \quad (BFGS step)
\Until{$\|\mathbf{U}^{(t)}\| < \varepsilon$ or maximum iterations reached}
\State $\hat{\bm{\Theta}}^\star \leftarrow \bm{\Theta}^{\star(t)}$; $\hat{\phi} \leftarrow \exp(\hat{\psi})$
\Ensure{$\hat{\bm{\gamma}}$, $\hat{\phi}$, $\mathbf{H}(\hat{\bm{\Theta}}^\star)$ (observed Hessian at MLE)}
\end{algorithmic}
\end{algorithm}

\section{Asymptotic Inference and Diagnostics}
\label{sec:inference}

\subsection{Asymptotic Covariance Matrix}

Under standard regularity conditions \cite{lehmann1998}, the MLE
$\hat{\bm{\Theta}}^\star$ is consistent and asymptotically normal:
\begin{equation}
  \sqrt{n}\,(\hat{\bm{\Theta}}^\star - \bm{\Theta}^{\star0})
  \xrightarrow{d} \mathcal{N}_{p+1}\!\left(\bm{0},\,
    \mathcal{I}(\bm{\Theta}^{\star0})^{-1}\right),
  \label{eq:asymp}
\end{equation}
where $\mathcal{I}(\bm{\Theta}^\star) = -\mathbb{E}[\nabla^2_{\bm{\Theta}^\star}\ell]$
is the expected Fisher information matrix. In practice, $\mathcal{I}$ is
estimated by the observed information:
\begin{equation}
  \widehat{\mathrm{Var}}(\hat{\bm{\Theta}}^\star)
  = \left[-\mathbf{H}(\hat{\bm{\Theta}}^\star)\right]^{-1},
  \label{eq:vcov}
\end{equation}
where $\mathbf{H}(\hat{\bm{\Theta}}^\star) = \nabla^2_{\bm{\Theta}^\star}\ell$
evaluated at the MLE is obtained from BFGS at convergence. Standard errors
for $\hat{\bm{\gamma}}$ are extracted from the upper-left $p \times p$
block of \eqref{eq:vcov}. Wald statistics and two-sided $p$-values follow
from the asymptotic normality of individual coefficient estimates.
Confidence intervals for $\hat{\phi}$ are constructed on the log scale and
back-transformed.

\subsection{Randomized Quantile Residuals}

Standard residuals defined via the Pearson or deviance paradigm are
unsuitable for modal regression because the response distribution is
reparameterized through $M_i$ rather than the mean. Instead, we adopt the
\emph{randomized quantile residuals} of \cite{dunn1996}. Define
the marginal CDF evaluated at the observed value:
\begin{equation}
  u_i = F(y_i \mid \hat{M}_i, \hat{\phi}), \quad i = 1, \ldots, n.
  \label{eq:ui}
\end{equation}
For right-censored observations ($c_i = 1$), the true event time
exceeds $y_i$, so $u_i$ is known only to lie in $[F(y_i \mid \hat{M}_i, \hat{\phi}), 1]$.
Randomization is applied:
\begin{equation}
  u_i \sim \mathrm{Uniform}\!\left[F(y_i \mid \hat{M}_i, \hat{\phi}),\, 1\right],
  \quad c_i = 1.
  \label{eq:rand}
\end{equation}
The randomized quantile residual is then $r_i = \Phi^{-1}(u_i)$,
where $\Phi^{-1}$ is the standard normal quantile function. Under the
correctly specified model, $r_i \overset{\mathrm{iid}}{\sim} \mathcal{N}(0,1)$
for all $i$, including censored units \cite{dunn1996}. This property
holds regardless of the response family, facilitating unified graphical
diagnostics: a normal Q-Q plot of $\{r_i\}$ should follow the reference
line, and a plot of $r_i$ against $\hat{M}_i$ should show no pattern.

\section{Simulation Study}
\label{sec:simulation}

A comprehensive Monte Carlo study with $B=1000$ replications is conducted to evaluate the finite-sample performance of the MLE across five distribution families. We consider sample sizes $n \in \{25, 50, 100, 200, 400\}$ and censoring proportions of $0\%$, $10\%$, and $25\%$. Performance is measured by empirical bias, root mean squared error (RMSE), and the coverage probability of 95\% Wald confidence intervals.

The data-generating process follows the modal regression framework with $\bm{\gamma} = (\gamma_0, \gamma_1, \gamma_2)^\top$ under three configurations:

\begin{itemize}
  \item \textbf{Gamma, Weibull, and Lognormal} (log link): $x_{1i}, x_{2i} \overset{\mathrm{iid}}{\sim} U(0,1)$ with $\bm{\gamma} = (0.8, 0.3, 0.15)^\top$.
  \item \textbf{Beta} (logit link): $x_{1i}, x_{2i} \overset{\mathrm{iid}}{\sim} U(-1,1)$ with $\bm{\gamma} = (-1.1, 0.3, 0.2)^\top$.
  \item \textbf{Inverse Gaussian} (log link): $x_{1i}, x_{2i} \overset{\mathrm{iid}}{\sim} U(0,1)$ with $\bm{\gamma} = (-2.0, 0.3, 0.15)^\top$. The intercept ensures the restriction $\lambda > 3M_i$ is satisfied, guaranteeing numerical stability.
\end{itemize}


\begin{landscape}
\begin{table}[!p]
  \caption{Empirical coverage probability (CP) of 95\% Wald intervals.}
  \label{tab:coverage_transposed}
  \centering
  \footnotesize
  \setlength{\tabcolsep}{4pt}
  \begin{tabular}{ll ccccc ccccc ccccc}
    \toprule
    & & \multicolumn{5}{c}{$p_c = 0\%$} & \multicolumn{5}{c}{$p_c = 10\%$} & \multicolumn{5}{c}{$p_c = 25\%$} \\
    \cmidrule(lr){3-7} \cmidrule(lr){8-12} \cmidrule(lr){13-17}
    \textbf{Family} & \textbf{Par.} & 25 & 50 & 100 & 200 & 400 & 25 & 50 & 100 & 200 & 400 & 25 & 50 & 100 & 200 & 400 \\
    \midrule
    \multirow{3}{*}{Lognormal} 
      & $\gamma_0$ & 0.916 & 0.930 & 0.943 & 0.945 & 0.947 & 0.917 & 0.932 & 0.937 & 0.934 & 0.936 & 0.912 & 0.921 & 0.941 & 0.957 & 0.951 \\
      & $\gamma_1$ & 0.913 & 0.939 & 0.946 & 0.938 & 0.954 & 0.927 & 0.930 & 0.950 & 0.939 & 0.951 & 0.924 & 0.938 & 0.936 & 0.955 & 0.949 \\
      & $\gamma_2$ & 0.923 & 0.933 & 0.950 & 0.963 & 0.953 & 0.914 & 0.940 & 0.941 & 0.950 & 0.954 & 0.917 & 0.920 & 0.945 & 0.946 & 0.960 \\
    \midrule
    \multirow{3}{*}{Gamma} 
      & $\gamma_0$ & 0.917 & 0.934 & 0.945 & 0.953 & 0.954 & 0.928 & 0.939 & 0.943 & 0.942 & 0.948 & 0.905 & 0.939 & 0.943 & 0.956 & 0.958 \\
      & $\gamma_1$ & 0.919 & 0.938 & 0.934 & 0.943 & 0.954 & 0.917 & 0.939 & 0.936 & 0.948 & 0.955 & 0.913 & 0.936 & 0.951 & 0.954 & 0.962 \\
      & $\gamma_2$ & 0.919 & 0.929 & 0.956 & 0.944 & 0.961 & 0.918 & 0.920 & 0.956 & 0.938 & 0.946 & 0.911 & 0.929 & 0.942 & 0.947 & 0.956 \\
    \midrule
    \multirow{3}{*}{Weibull} 
      & $\gamma_0$ & 0.912 & 0.922 & 0.945 & 0.946 & 0.944 & 0.899 & 0.935 & 0.950 & 0.945 & 0.938 & 0.904 & 0.948 & 0.945 & 0.928 & 0.936 \\
      & $\gamma_1$ & 0.923 & 0.941 & 0.953 & 0.938 & 0.951 & 0.899 & 0.939 & 0.939 & 0.947 & 0.937 & 0.909 & 0.937 & 0.937 & 0.946 & 0.961 \\
      & $\gamma_2$ & 0.915 & 0.934 & 0.959 & 0.938 & 0.957 & 0.905 & 0.935 & 0.947 & 0.949 & 0.939 & 0.903 & 0.936 & 0.938 & 0.937 & 0.942 \\
    \midrule
    \multirow{3}{*}{Beta} 
      & $\gamma_0$ & 0.927 & 0.926 & 0.943 & 0.944 & 0.942 & 0.908 & 0.938 & 0.953 & 0.936 & 0.960 & 0.904 & 0.925 & 0.937 & 0.954 & 0.953 \\
      & $\gamma_1$ & 0.915 & 0.945 & 0.931 & 0.950 & 0.934 & 0.916 & 0.948 & 0.952 & 0.939 & 0.963 & 0.929 & 0.928 & 0.930 & 0.959 & 0.946 \\
      & $\gamma_2$ & 0.915 & 0.934 & 0.947 & 0.942 & 0.948 & 0.920 & 0.953 & 0.936 & 0.945 & 0.952 & 0.923 & 0.929 & 0.941 & 0.945 & 0.965 \\
    \midrule
    \multirow{3}{*}{Inv. Gaussian} 
      & $\gamma_0$ & 0.909 & 0.949 & 0.936 & 0.929 & 0.953 & 0.916 & 0.929 & 0.952 & 0.922 & 0.947 & 0.894 & 0.922 & 0.954 & 0.949 & 0.949 \\
      & $\gamma_1$ & 0.917 & 0.947 & 0.946 & 0.950 & 0.954 & 0.919 & 0.929 & 0.941 & 0.945 & 0.945 & 0.900 & 0.950 & 0.942 & 0.943 & 0.956 \\
      & $\gamma_2$ & 0.920 & 0.942 & 0.949 & 0.955 & 0.947 & 0.896 & 0.940 & 0.938 & 0.949 & 0.947 & 0.896 & 0.924 & 0.948 & 0.951 & 0.941 \\
    \bottomrule
  \end{tabular}
\end{table}
\end{landscape}

The internal dispersion parameters are fixed at $\alpha = 3.5$ (Gamma), $k = 2.5$ (Weibull), $\alpha = 3.0$ (Beta), $\sigma^2 = 0.5$ (Lognormal), and $\lambda = 5.0$ (Inverse Gaussian).

Right censoring is generated independently of the event times.
For a target censoring proportion $p_c \in \{0\%, 10\%, 25\%\}$,
the censoring time is drawn from an exponential distribution
$W_i \sim \mathrm{Exp}(\lambda_c)$, where $\lambda_c$ is calibrated
via a preliminary pilot run of 50\,000 observations so that the
empirical censoring proportion approximates the target.
When $p_c = 0\%$, no censoring is applied.

The sample size varies over $n \in \{25, 50, 100, 200, 400\}$, giving $15$ scenarios per distribution family and $75$ scenarios in total. For each scenario, $B = 1{,}000$ Monte Carlo replications are run. In each replication $b = 1, \ldots, B$:

\begin{enumerate}
  \item Generate covariates $\{(x_{1i}, x_{2i})\}_{i=1}^n$ and compute
    the true conditional mode $M_i$ via the inverse link.
  \item Draw event times $Y_i$ from the family-specific distribution
    parameterized through $M_i$ and the true dispersion.
  \item Draw censoring times $W_i$ and form the observed pair
    $(y_i, c_i) = (\min(Y_i, W_i),\, \mathbf{1}\{Y_i > W_i\})$.
  \item Fit the modal regression model via
    \texttt{modal\_cens(y \textasciitilde\ x1 + x2, data, cens, family)}
    using Algorithm~\ref{alg:mle}.
  \item Record $\hat{\bm{\gamma}}^{(b)}$, $\hat{\phi}^{(b)}$,
    and the Wald 95\% confidence interval for each $\gamma_j$.
\end{enumerate}

\begin{figure}[!b]
  \centering
  \includegraphics[width=0.85\textwidth]{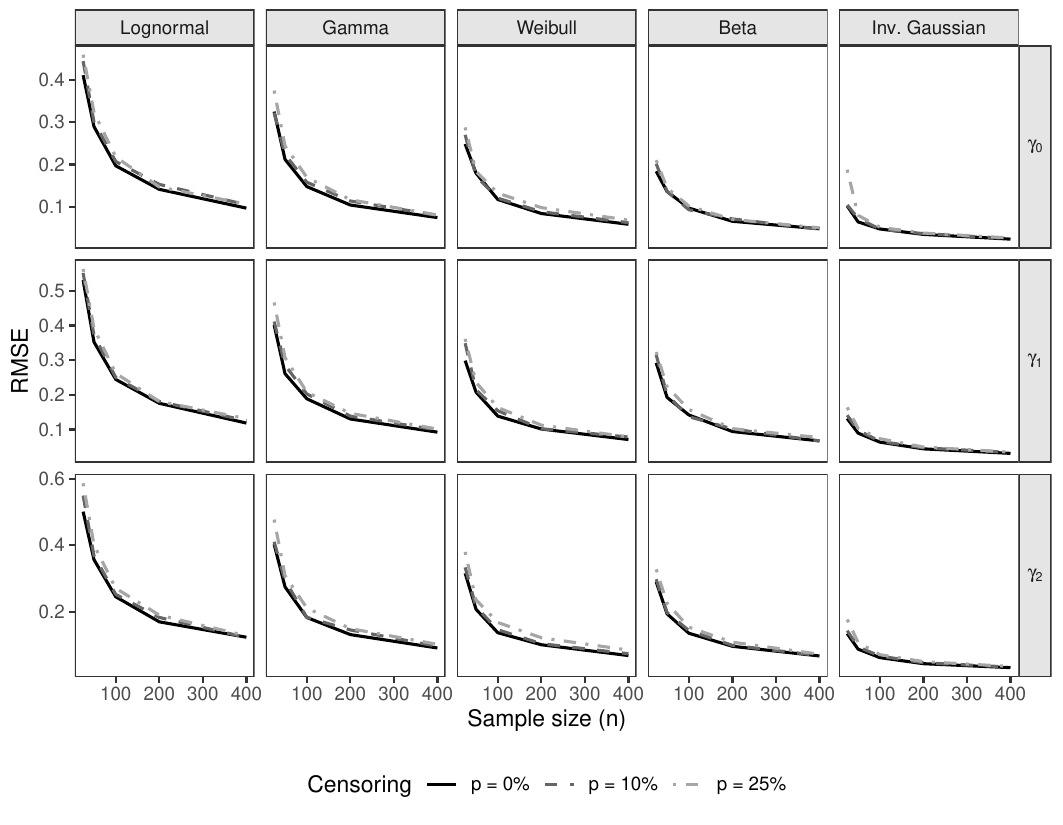}
  \caption{Root mean squared error (RMSE) of the regression coefficient
    estimates across $B = 1000$ replications. Rows: regression
    parameters ($\gamma_0$, $\gamma_1$, $\gamma_2$). Columns:
    distribution families. Line types and shades distinguish the
    censoring levels.}
  \label{fig:rmse_lines}
\end{figure}

For each parameter $\gamma_j$ ($j = 0, 1, 2$), the following metrics
are computed over the $B$ successful replications:

\begin{itemize}
  \item \textbf{Empirical Bias:}
    $\mathrm{Bias}(\hat\gamma_j) = \bar{\hat\gamma}_j - \gamma_j^0$,
    where $\bar{\hat\gamma}_j = B^{-1}\sum_{b=1}^B \hat\gamma_j^{(b)}$.
  \item \textbf{Root Mean Squared Error:}
    $\mathrm{RMSE}(\hat\gamma_j) = \left[B^{-1}\sum_{b=1}^B
    (\hat\gamma_j^{(b)} - \gamma_j^0)^2\right]^{1/2}$.
  \item \textbf{Empirical Coverage Probability (CP):}
    the proportion of the $B$ replications in which the true
    $\gamma_j^0$ falls within the 95\% Wald interval
    $\hat\gamma_j^{(b)} \pm 1.96\,\widehat{\mathrm{SE}}_j^{(b)}$.
\end{itemize}

\noindent Analogous metrics are computed for the dispersion parameter
$\phi$ on its original scale.

Figure~\ref{fig:rmse_lines} summarizes the root mean squared error
(RMSE) of the regression coefficient estimates across the five
distribution families. Each panel displays the RMSE as a function of
sample size, with separate curves for each censoring level. 
Detailed boxplots of the estimation error
$\hat\gamma_j^{(b)} - \gamma_j^0$ for each family are provided in
Appendix~\ref{app:boxplots}.


The simulation results confirm the theoretical properties of the MLE across all five families, as shown by the RMSE curves in Figure~\ref{fig:rmse_lines}, which decrease monotonically with sample size at a rate compatible with the expected $O(n^{-1/2})$ asymptotics. Although estimation errors are substantial at $n=25$ (especially for the intercept $\gamma_0$), the RMSE falls below $0.05$ for the majority of regression parameters by $n=400$ under every censoring scenario, indicating precise estimation even with 25\% right censoring. The effect of censoring is visible as a moderate upward shift of the RMSE curves due to information loss, but the convergence rate remains unaffected, confirming that the censored likelihood formulation preserves consistency. Notably, the Lognormal, Gamma, and Weibull families exhibit nearly identical trajectories for the slope parameters, while the Beta and Inverse Gaussian models also demonstrate robust recovery despite their specific link functions and parametric constraints.

Table~\ref{tab:coverage_transposed} reports the empirical coverage probability (CP) of the 95\% Wald confidence intervals for each regression coefficient. The nominal level is 95\%; under correct asymptotic theory, the CP should approach 0.95 as $n$ grows. Systematic under-coverage at small $n$ indicates that the normal approximation to the MLE sampling distribution has not yet become accurate, while over-coverage may suggest conservative standard errors.

\section{Application: Motorette Insulation Life Data}
\label{sec:application}

To illustrate the methodology on a real-world reliability problem, we analyze the accelerated life test data on electrical insulation in 40~motorettes reported by \cite{SchmeeHahn1979}. Ten motorettes were tested at each of four elevated temperatures ($150\,^{\circ}$C, $170\,^{\circ}$C, $190\,^{\circ}$C, and $220\,^{\circ}$C), and testing was terminated at different times at each temperature level. Of the 40~units, only $r = 17$ exhibited observed failures; the remaining $m - r = 23$ were right-censored, yielding a substantial censoring proportion of 57.5\%.

Following the Arrhenius transformation used in the original study, we define the response as $y_i = \log_{10}(\text{failure time}_i)$ and the covariate as $T_i = 1000 / (\text{Temperature}_i + 273.2)$, so that the linear predictor takes the form
\[
  \log(M_i) \;=\; \gamma_0 + \gamma_1 \, T_i,
\]
where $M_i$ denotes the conditional mode of the response given $T_i$.

We fit the modal regression model under all four positive continuous families supported by \texttt{ModalCens}: Gamma, Weibull, Lognormal, and Inverse Gaussian. All families employ a logarithmic link function for the mode, and the censoring mechanism is explicitly incorporated into the likelihood as described in Section~\ref{sec:cases}. The results for all four models are summarized in Table~\ref{tab:app_comparison}.

\begin{table}[!h]
  \caption{Application to the motorette data: comparison of modal regression models under the Weibull, Gamma, Lognormal, and Inverse Gaussian families. The best model according to AIC is indicated in bold.}
  \label{tab:app_comparison}
  \centering
  \begin{tabular}{lccccc}
    \toprule
    & \textbf{Weibull} & \textbf{Gamma} & \textbf{Lognormal} & \textbf{Inv. Gaussian} \\
    \midrule
    $\hat{\gamma}_0$ (Intercept) & $-1.6604$ & $-1.7263$ & $-1.7369$ & $-1.2598$ \\[3pt]
    $\hat{\gamma}_1$ ($T_i$)     & $\phantom{-}1.3194$  & $\phantom{-}1.3445$  & $\phantom{-}1.3485$ & $\phantom{-}0.4578$ \\
    \midrule
    $\ell(\hat{\boldsymbol{\theta}})$ & $-9.746$ & $-12.076$ & $-12.303$ & $-62.292$ \\[2pt]
    AIC  & $\phantom{-}\mathbf{25.49}$ & $\phantom{-}30.15$ & $\phantom{-}30.61$ & $\phantom{-}130.58$ \\[2pt]
    BIC  & \phantom{-}$\mathbf{30.56}$ & $\phantom{-}35.22$ & $\phantom{-}35.67$ & $\phantom{-}135.65$ \\
    \bottomrule
  \end{tabular}
\end{table}

Among the families, Weibull yields the lowest AIC ($25.49$) and BIC ($30.56$). Gamma and Lognormal provide competitive fits, while Inverse Gaussian performs poorly (AIC = $130.58$). This lack of fit stems from the restriction $\lambda > 3M_i$, which forces the estimated mode significantly below the observed range. Conversely, the three competitive models produce stable estimates for $\gamma_1$ ($1.319$--$1.349$) and $\gamma_0$ ($-1.737$ to $-1.660$), demonstrating the robustness of the modal approach.With a logarithmic link, $\gamma_1$ is directly interpretable: a unit increase in $T_i$ multiplies the modal failure time by approximately $\exp(1.319) \approx 3.74$. This aligns with the physical expectation that lower temperatures extend insulation life.

\begin{figure}[!t]\centering\includegraphics[width=\textwidth]{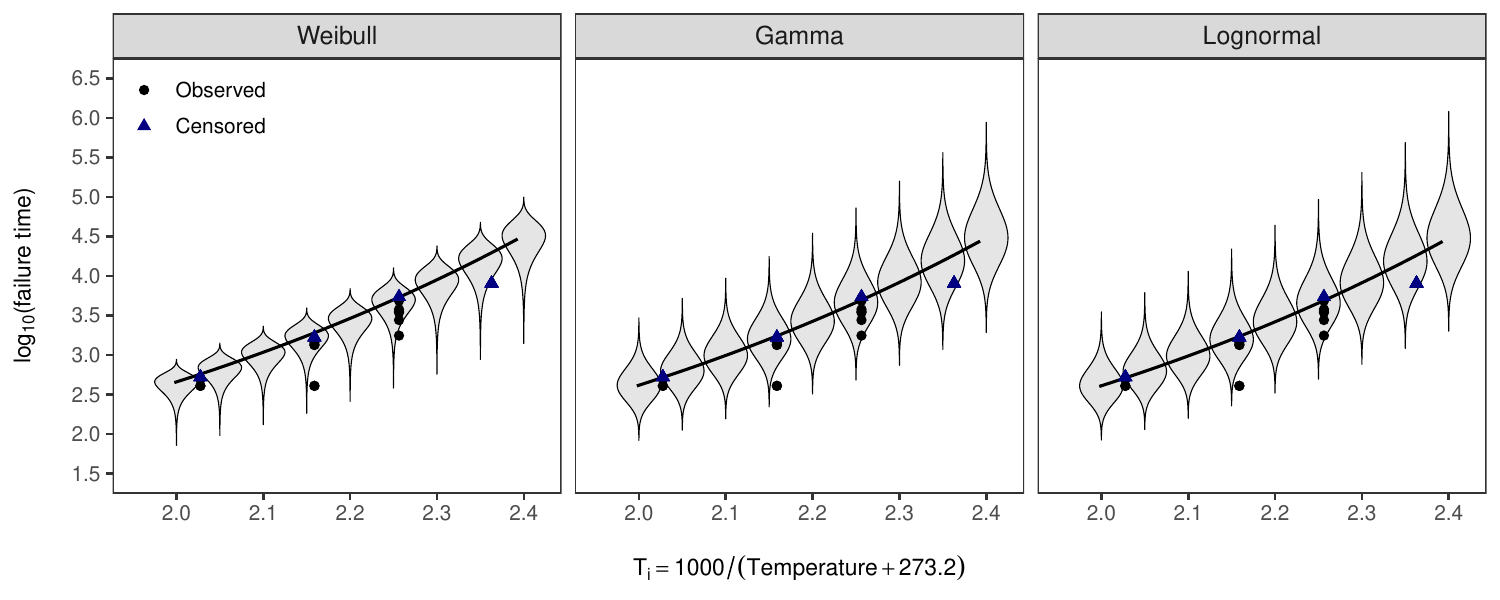}
\caption{Conditional densities and fitted modal regression lines (Weibull, Gamma, and Lognormal). Inverse Gaussian is excluded due to poor fit from the $\lambda > 3M_i$ restriction. Circles: observed failures; triangles: right-censored observations.}
\label{fig:app_fitted}
\end{figure}

Figure~\ref{fig:app_fitted} illustrates the modal lines and conditional densities for the Weibull, Gamma, and Lognormal models. Inverse Gaussian is omitted as its restricted trajectory fails to follow the data. The visualization confirms that the regression lines accurately track the conditional modes, with Weibull best capturing the central tendency under heavy censoring.

\begin{figure}[!h]
  \centering
  \includegraphics[width=0.9\textwidth]{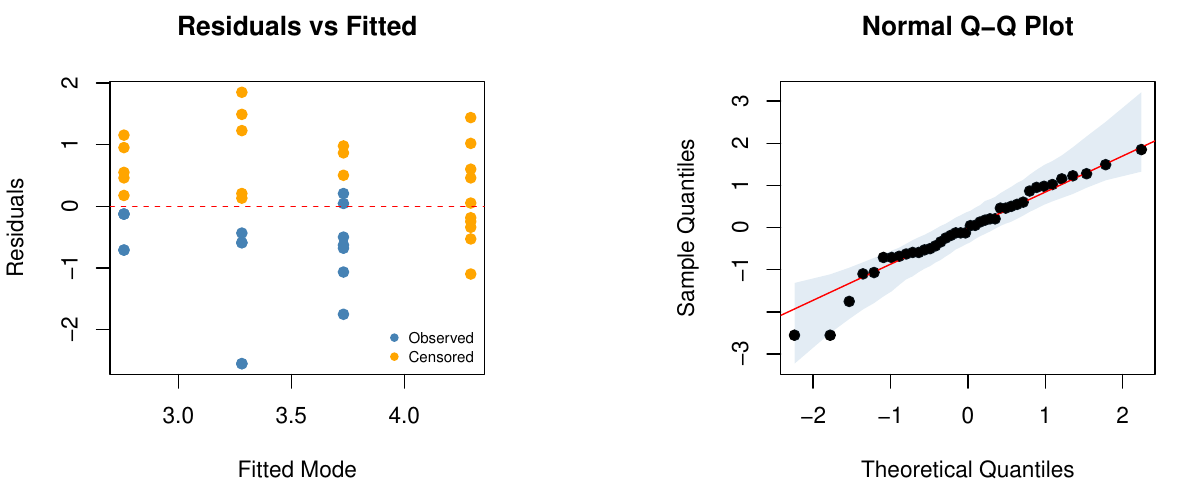}
  \caption{Diagnostic plots for the Weibull modal regression on the motorette data. Left: randomized quantile residuals versus fitted modes. Right: normal Q-Q plot.}
  \label{fig:app_residuals}
\end{figure}

Furthermore, Figure~\ref{fig:app_residuals} displays diagnostics for the Weibull model. The randomized quantile residuals and the normal Q-Q plot support the model's adequacy, showing no systematic patterns despite the substantial censoring.

\section{Conclusions and Future Work}
\label{sec:conclusion}

We presented a self-contained parametric framework for modal regression applicable to positive distributions under right censoring. By analytically mapping the distributional parameters directly to the conditional mode, we transformed the estimation into a standard maximum likelihood problem. Asymptotic inference and randomized quantile residuals complete the proposed methodology, providing robust estimation and diagnostics.

The framework is fully implemented in the \texttt{ModalCens} R package. Extensive Monte Carlo simulations confirmed the consistency of the estimators, showing that bias and RMSE decrease at the expected rate and that Wald confidence intervals achieve nominal coverage even under heavy censoring. Finally, the practical utility of our approach was successfully demonstrated through a real-world application to accelerated life test data, highlighting its robustness and interpretability in applied reliability analysis.

Several directions for future research naturally emerge from the proposed framework. Developing a fully Bayesian approach for modal regression could provide robust finite-sample inference without relying on asymptotic normality. Additionally, integrating penalized likelihood techniques would extend this methodology to allow for simultaneous estimation and variable selection in high-dimensional settings.

\clearpage

\appendix
\section{Estimation Error Boxplots}
\label{app:boxplots}

\begin{figure}[!h]
  \centering
  \includegraphics[width=0.8\textwidth]{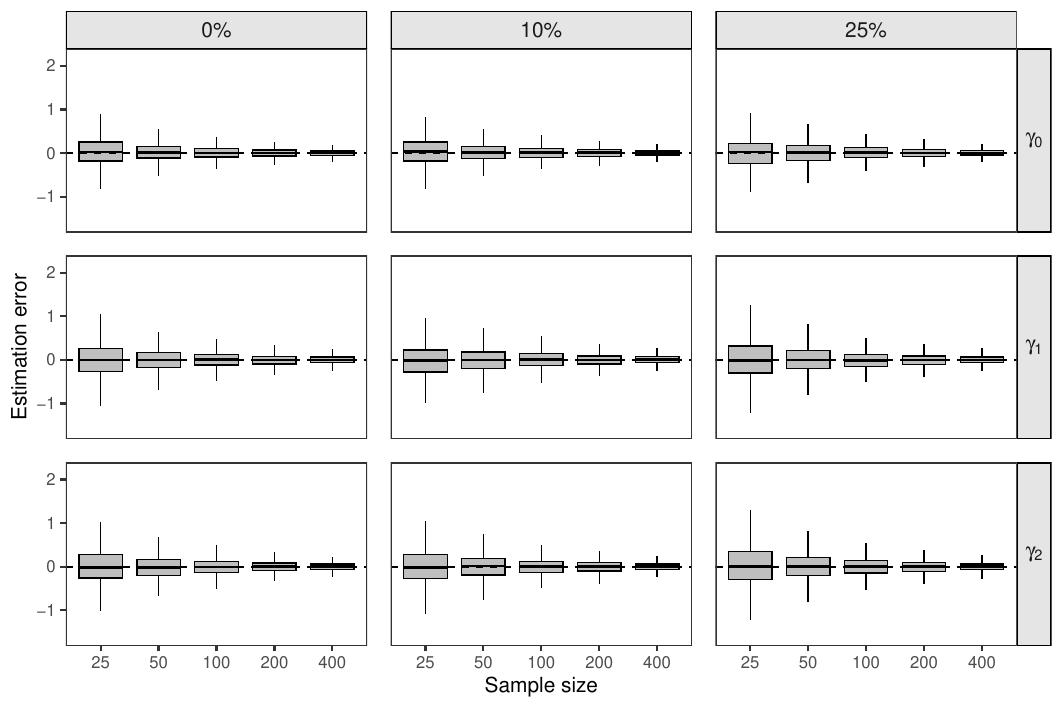}
  \caption{Gamma family: boxplots of the estimation error
    $\hat\gamma_j - \gamma_j^0$ across $B = 1{,}000$ replications.}
  \label{fig:rmse_gamma}
\end{figure}

\begin{figure}[!h]
  \centering
  \includegraphics[width=0.8\textwidth]{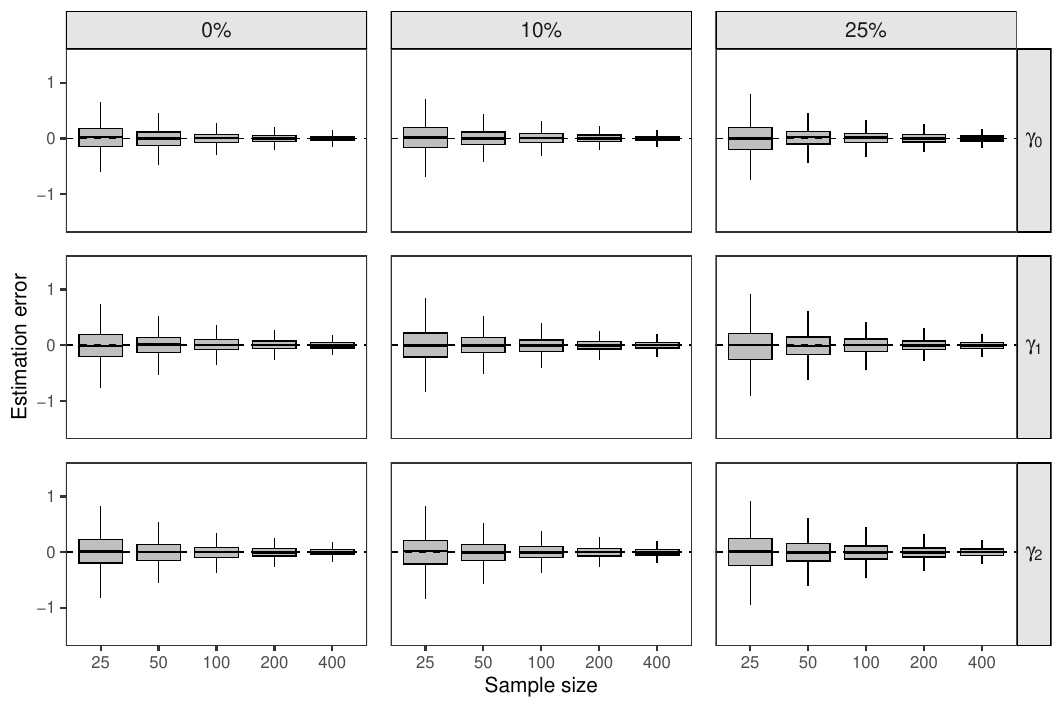}
  \caption{Weibull family: boxplots of the estimation error
    $\hat\gamma_j - \gamma_j^0$ across $B = 1{,}000$ replications.}
  \label{fig:rmse_weibull}
\end{figure}

\begin{figure}[!h]
  \centering
  \includegraphics[width=0.8\textwidth]{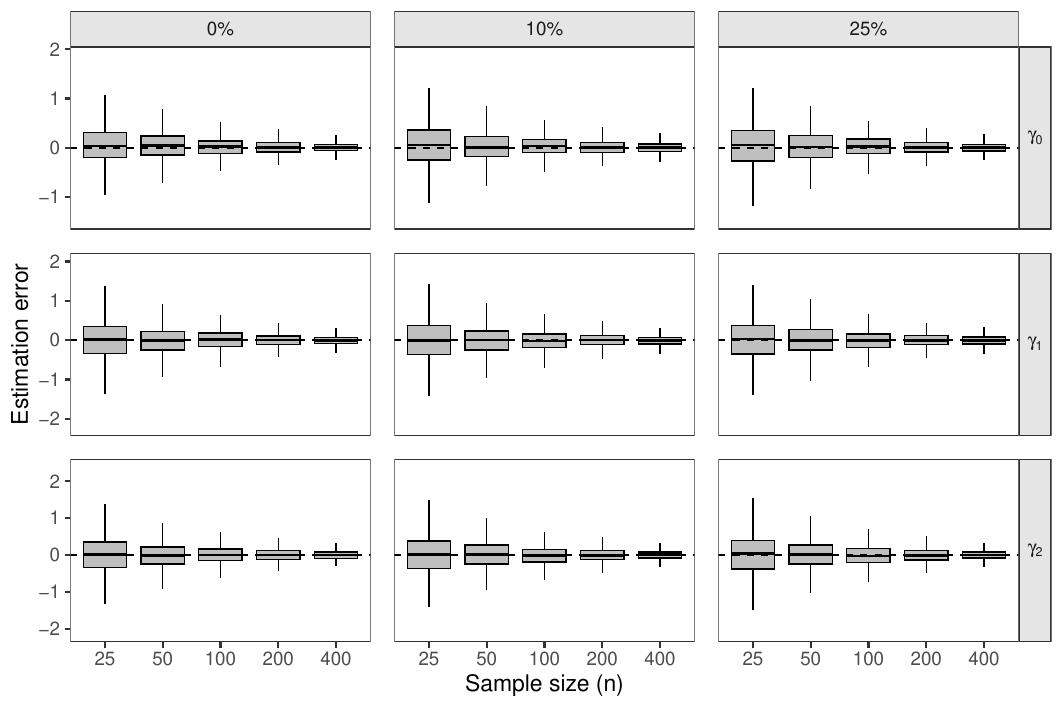}
  \caption{Lognormal family: boxplots of the estimation error
    $\hat\gamma_j - \gamma_j^0$ across $B = 1{,}000$ replications.}
  \label{fig:rmse_lognormal}
\end{figure}

\begin{figure}[!h]
  \centering
  \includegraphics[width=0.8\textwidth]{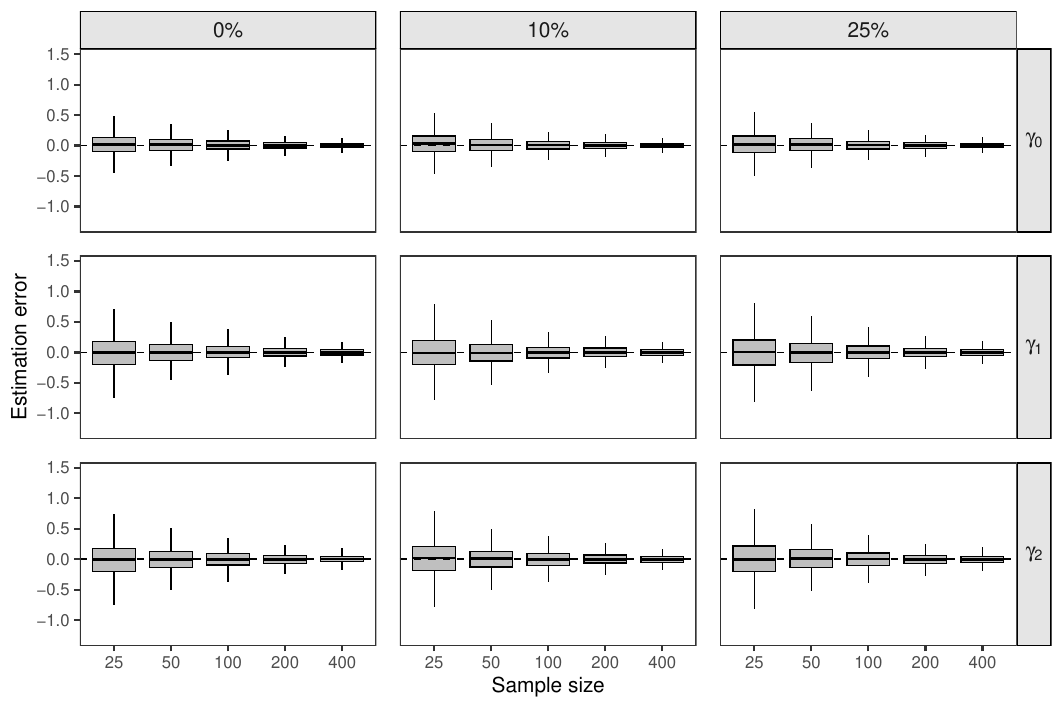}
  \caption{Beta family: boxplots of the estimation error
    $\hat\gamma_j - \gamma_j^0$ across $B = 1{,}000$ replications.}
  \label{fig:rmse_beta}
\end{figure}

\begin{figure}[!h]
  \centering
  \includegraphics[width=0.8\textwidth]{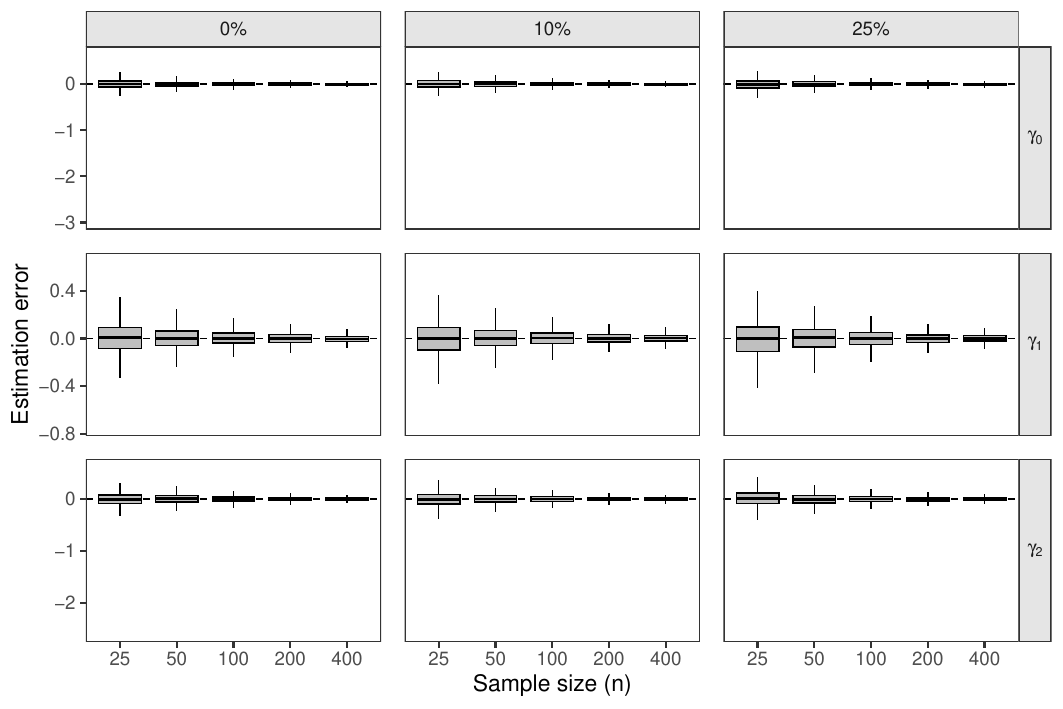}
  \caption{Inverse Gaussian family: boxplots of the estimation error
    $\hat\gamma_j - \gamma_j^0$ across $B = 1{,}000$ replications.}
  \label{fig:rmse_ig}
\end{figure}

Figures~\ref{fig:rmse_gamma}--\ref{fig:rmse_ig} provide detailed
boxplots of the estimation error $\hat\gamma_j^{(b)} - \gamma_j^0$
across the $B = 1{,}000$ Monte Carlo replications for each regression
coefficient. Each figure corresponds to a single distribution family
and is organized as a $3 \times 3$ facet grid with rows corresponding
to the three regression parameters ($\gamma_0$, $\gamma_1$, $\gamma_2$)
and columns to the three censoring levels ($0\%$, $10\%$, $25\%$).

\vspace{5cm}

\section{The \texttt{ModalCens} R Package}
\label{sec:package}

\texttt{ModalCens} is an open-source R package implementing all methodology
described in Sections~\ref{sec:framework}--\ref{sec:inference}. It is
available at \url{https://github.com/chedgala/ModalCens} and installed via:

\vspace{4cm}

\begin{lstlisting}
# install.packages("devtools")
devtools::install_github("chedgala/ModalCens")
\end{lstlisting}

The package exports a single primary function, \texttt{modal\_cens()},
together with a complete set of S3 methods for the returned object of
class \texttt{"ModalCens"}.
\medskip
\begin{lstlisting}
modal_cens(formula, data, cens, family = "gamma")
\end{lstlisting}

\noindent\textbf{Arguments:}
\begin{itemize}
  \item \texttt{formula}: an R formula, e.g., \texttt{y \textasciitilde\ x1 + x2}.
  \item \texttt{data}: a data frame; the function halts with an informative
        error if any variable referenced by \texttt{formula} contains
        \texttt{NA} values (no imputation is performed).
  \item \texttt{cens}: a binary numeric vector with \texttt{1} = right-censored,
        \texttt{0} = fully observed; must be free of \texttt{NA}s and have the
        same length as \texttt{nrow(data)}.
  \item \texttt{family}: one of \texttt{"gamma"}, \texttt{"beta"},
        \texttt{"weibull"}, \texttt{"lognormal"}, \texttt{"invgauss"}.
\end{itemize}

\noindent\textbf{Return value:} an object of class \texttt{"ModalCens"} with
components listed in Table~\ref{tab:return}.

\begin{table}[b]
  \caption{Components of a \texttt{"ModalCens"} object.}
  \label{tab:return}
  \centering
  \small
  \begin{tabular}{lp{9cm}}
    \toprule
    Component & Description \\
    \midrule
    \texttt{coefficients} & $\hat{\bm{\gamma}}$: estimated regression coefficients. \\
    \texttt{phi} & $\hat{\phi}$: estimated dispersion/shape (original scale). \\
    \texttt{vcov} & Full $(p+1)\times(p+1)$ asymptotic covariance matrix (includes \texttt{log\_phi}). \\
    \texttt{vcov\_beta} & $p\times p$ sub-matrix for regression coefficients only. \\
    \texttt{fitted.values} & Estimated conditional modes $\hat{M}_i$. \\
    \texttt{residuals} & Randomized quantile residuals $r_i$ (Dunn--Smyth). \\
    \texttt{loglik} & Maximized log-likelihood $\hat{\ell}$. \\
    \texttt{n} & Number of observations. \\
    \texttt{n\_par} & Total number of estimated parameters ($p+1$). \\
    \texttt{family} & Distribution family used. \\
    \texttt{cens} & Censoring indicator as supplied. \\
    \bottomrule
  \end{tabular}
\end{table}

Table~\ref{tab:methods} summarizes all generic methods dispatched on \texttt{"ModalCens"} objects. Internally, model optimization relies on the BFGS algorithm with analytical Hessians, applying a strict tolerance (\texttt{reltol = 1e-10}) and a 2000-iteration limit. Initial values for $\bm{\gamma}$ are derived via ordinary least squares on the link-transformed response, while the dispersion parameter is estimated on the log scale ($\psi = \log\phi$) to ensure positivity. For the Inverse Gaussian family, a smooth penalty on the negative log-likelihood enforces the $\lambda > 3M_i$ constraint while preserving gradient continuity. Finally, numerical stability for censored randomized quantile residuals is maintained by clamping the uniform draws on $[F(y_i), 1]$ to the $[10^{-7}, 1-10^{-7}]$ interval.

\begin{table}[!t]
  \caption{S3 methods for objects of class \texttt{"ModalCens"}.}
  \label{tab:methods}
  \centering
  \begin{tabular}{lp{9.5cm}}
    \toprule
    Method & Description \\
    \midrule
    \texttt{summary()} & Coefficient table with SE, Wald $z$-statistic, and two-sided $p$-value;
                          $\hat{\phi}$; maximized log-likelihood; AIC; BIC; pseudo-$R^2$,
                          computed as the squared Pearson correlation between observed
                          responses and fitted modes, $R^2 = [\mathrm{cor}(y_i, \hat{M}_i)]^2$. \\
    \texttt{plot()} & Two-panel diagnostic: (i) residuals vs.\ fitted modes, distinguishing
                       observed and censored points; (ii) normal Q-Q plot of $r_i$. \\
    \texttt{coef()} & $\hat{\bm{\gamma}}$. \\
    \texttt{vcov()} & Full $(p+1)\times(p+1)$ covariance matrix. \\
    \texttt{residuals()} & Randomized quantile residuals. \\
    \texttt{logLik()} & Maximized log-likelihood. \\
    \texttt{AIC()} / \texttt{BIC()} & Information criteria: $-2\hat{\ell} + k(p+1)$, $k=2$ or $k=\log n$. \\
    \texttt{confint()} & Wald confidence intervals for $\hat{\bm{\gamma}}$. \\
    \bottomrule
  \end{tabular}
\end{table}

\end{document}